\documentclass[12pt]{iopart}
\usepackage{graphicx}
\begin{document}

\title[Soliton ejection from low-power probes]{Induced soliton
ejection from a continuous-wave source waveguided by an optical pulse-soliton train}

\author{Alain~M.~Dikand\'e}

\address{Laboratory of Research on Advanced Materials and Nonlinear
Sciences~(LaRAMaNS), Department of Physics, Faculty of Science, University of
Buea P.O. Box 63 Buea, Cameroon}
\ead{adikande@ictp.it}
\begin{abstract}
It has been established for some time that high-power pump can trap a probe beam of lower intensity that is simultaneously
propagating in a Kerr-type optical medium, inducing a focusing of the probe with
the 
emergence of modes displaying solitonic properties. To understand the
mechanism by which such self-sustained modes are generated, and mainly the 
changes on probe spectrum induced by the cross-phase-modulation effect for
an harmonic probe trapped by a multiplex of temporal pulses, a linear 
equation (for the probe) and a nonlinear Schr\"odinger equation (for the pump)
both coupled by a cross-phase-modulation term, are considered 
simultaneously. In general the set of coupled probe-pump equations is not 
exactly tractable at any arbitrary value of the ratio of the cross-phase to the
self-phase modulation strengths. However, for certain values of this 
ratio, the probe modulation wavector develops into $\vert n,l\textgreater$ {\it
quantum states} involving $2n+1$ soliton-shaped eigenfunctions which spectral 
properties can be characterized unambiguously. Solutions of the probe equation
give evidence that the competition between the self-phase and cross-phase 
modulations leads to a broadband spectrum, with the possibility of a
quasi-continuum of soliton modes when the cross-phase-modulation coupling is
strong enough.

\end{abstract}

\vspace{2pc}
\submitto{\JOA}
\maketitle

\section{Introduction}
In two classic papers~\cite{manas1, manas2}, by solving the master-slave optical system
\begin{eqnarray}
-i\frac{\partial A_1}{\partial z} - \frac{1}{2k_1}\frac{\partial^2 A_1}{\partial x^2}&=&\frac{n_2^{spm}k_1}{n_0}\vert A_1\vert^2 A_1, \label{manas1} \\
-i\frac{\partial A_2}{\partial z} - \frac{1}{2k_2}\frac{\partial^2 A_2}{\partial x^2}&=&\frac{n_2^{ipm}k_2}{n_0}\vert A_1\vert^2 A_2, \label{manas2}
\end{eqnarray}
the author predicted that ultrafast pulses with energies lower than that required to self-sustain soliton shape in
the anomalous dispersion regime of an optical medium may preserve their shape,
provided an intense copropagating pump of a different color (i.e. wavelength)
and a longer duration typical of soliton is present. Since this key input as
well as the subsequent experimental evidences reported in~\cite{fuente1,fuente2}
it has become evident that besides their most common applications in
long-distance transmission of high-power signals~\cite{stolen,agarwal}, solitons
could be prime candidates for controlling light by light~\cite{elena,malom}.
More advanced studies have been carried out in recent years based on numerical
simulations, and revealed that due to their unique stability properties solitons
provide excellent and reliable guiding structures for reconfigurable low-power
signals and re-routinable high-power pulses~\cite{rand2,rand3}. Particularly
attracting is the possibility to construct soliton networks that can be used as
nonlinear waveguide arrays, as established in the landscape of recent
experiments~\cite{6,8,9,10,11,adolfo} emphasizing the outstanding robustness of
the periodic lattices of optical solitons with flexible (hence controllable)
refractive index modulation depth and period that are induced all-optically. In
general, these periodic optical-soliton structures form in continuous nonlinear
media where they develop into optically imprinted modulations with some
effective refractive index. Given the tunable character of the effective
refractive index as well as the flexible modulation period, a great number of
new opportunities for all-optical manipulation of light can be
envisaged since in this case, periodic optical-soliton waveguides can operate in
both weak and strong-coupling regimes depending on the depth of refractive index
modulation. Transmission techniques exploiting optically-induced waveguiding
configurations with solitons are now common in several communication media such
as laser~\cite{naumov}, photon~\cite{rand2,rand3,dika1} systems and
photorefractive semiconductor~\cite{chen} media. \\
The present work aims at extending the phenomenon of signal reconfiguration by
strong optical fields to the context of an harmonic cw beam trapped in the
guiding structure of a periodic pulse train. As in a previous study~\cite{dika1}
we assume that due to the competition between the self-phase modulation (SPM)
and cross-phase modulation (XPM)~\cite{naumov,mol1,shapiro1,shapiro2,segev4}
effects, the low-power probe can be reshaped giving rise to self-sustained
optical signals trapped within the guiding structure of the temporal pulse
multiplex. But before it is relavant to stress that the groundstate spectrum of
the trapped probe in the case of one-soliton pump has been discussed in details~\cite{rand2,rand3}. Thus, it is established both
analytically and numerically that even in the steady-state regime of the pump
propagation the probe groundstate is not well defined at any arbitrary value of
the ratio between the cross-phase and self-phase modulation strengths but only
for some specific choices of this ratio. For one particular value of this ratio involving well defined optical structures,
it has been found~\cite{manas1,manas2,rand2} that the fundamental mode of the trapped probe groundstate consists of a single-pulse soliton
with finite momentum. Very recently, considering the same
particular value of the XPM to the SPM ratio we pointed out~\cite{dika1} that
the physics of the pump-probe system turns to be quite rich if the pump is a
periodic train of pulses, time-multiplexed at the input of an optical medium as
for instance an optical fiber prior to propagation. Namely, we found that in
response to this temporal pulse multiplexing in the pump the trapped probe
spectrum could burst into a band reflecting distinct possible induced fundamental soliton modes.
\\
Still the competition between the SPM and XPM effects is never of the same
order and can change from one optical medium to another, implying quite distinct
features and properties of the probe groundstate. In this last respect, for the
value of the XPM to the SPM ratio considered in our previous study we
found~\cite{dika1} three distinct soliton modes, and observed that they form a
complete orthogonal set of $\vert n,l\textgreater$ eigenstates two of which were
almost degenerate. \\
In the present study we shall go beyond previous considerations in terms of the
order of the competition between the XPM and SPM
effects~\cite{rand2,rand3,dika1}, by considering two new representative
but larger values of the ratio of their strengths. Our ultimate goal through
such assumptions is to point out an increasingly broadband and highly degenerate
spectrum for the induced-soliton states in the probe, so broad and degenerate
that the probe field can develop into a soliton quasi-continuum for sufficiently
strong XPM effect relative to the SPM effect.
\section{Probe source spectral problem} 
The pump-probe system of our current interest is described by the two coupled
equations~\cite{manas1,manas2,rand2,dika1}: 
\begin{eqnarray}
i\frac{\partial A_1}{\partial z} &+& \frac{1}{2}\frac{\partial^2 A_1}{\partial t^2}
+ \xi\vert A_1 \vert^2\,A_1= 0,
\label{a1a} \\
i\frac{\partial A_2}{\partial z} &+& \frac{1}{2}D\frac{\partial^2 A_2}{\partial t^2}
+ i\lambda\frac{\partial A_2}{\partial t} + \beta\vert A_1 \vert^2\,A_2 =0,
\label{a1b}
\end{eqnarray}
where~(\ref{a1a}) is the pump equation assumed to describe propagation in a
Kerr-type optical fiber in the anomalous dispersion
regime~\cite{manas1,manas2}, and~(\ref{a1b}) is the linear equation
corresponding to the harmonic probe. The quantities $A_1$ and $A_2$ are envelopes of
the pump and probe respectively, $D$ is the group-velocity dispersion of the
probe (here taken arbitrary to account for possible difference with that of the
pump), $\xi$ is the SPM coefficient generic of nonlinearity in the pump source,
$\lambda$ in the second equation is the (temporal) walk-off between the pump and
probe while $\beta$ measures the strength of the cross-phase interaction between
the pump and probe. \\
Our main point of focus is the probe equation~(\ref{a1b}) which, as it stands,
requires an explicit knowledge of shape profile of the fundamental mode composing the pump signal. As we are
interested in a pump consisting of a train of time-multiplexed pulses, we focus
on a classic implementation where the separation between pulses in the soliton
train is varied by controlling their mutual interactions, namely via the
dispersion-management technique (see more detailed discussions e.g.
in~\cite{menyuk1,menyuk3,duce}). It is known that such a periodic structure of
equally separated pulses can well be represented by the sum: 
\begin{equation}
\vert A_1(z_0, t)\vert= \sum_{n=1}^N{\vert Q_n\vert \,sech\,Q_n\left(t - t_n -
\omega_n z_0\right)} \label{a3}
\end{equation}
where, following parameter definitions in~\cite{menyuk3}, the quantity $N$
represents the number of channels in the time domain, $t_n$ is the initial
temporal position of a given $n^{th}$ pulse in the input multiplex while $Q_n$
and $\omega_n$ are the associate amplitude and central frequency respectively.
The input time-multiplexed pulse structure~(\ref{a3}) can be reduced to the
following single-valued function, using the exact summation rule for $sech$
functions~\cite{hansen,magnus}:
\begin{equation}
A_1(z, t)= \frac{Q}{\sqrt{\xi}}\, dn\left[Q\left(t-t_0 - \omega
z\right),\kappa\right]e^{i\left[\omega t + \frac{\kappa^2\omega^2}{2} z +
\frac{Q^2 - 2\omega^2}{2} z\right]}, \label{a2}
\end{equation}
which quite remarkably, coincides with the exact periodic-soliton solution of
the pump equation~(\ref{a1a})~\cite{karta3,karta5}. It is instructive specifying
that $dn$ in the above formula is the Jacobi elliptic function of modulus
$\kappa$, $Q$ is the amplitude and $\omega$ is the characteristic pulse
frequency. The Jacobi elliptic function $dn$ is periodic in its arguments $(z,
t)$ and for the solution~(\ref{a2}), the temporal period is $\tau_0=
2K(\kappa)/Q$ where $K(\kappa)$ is the elliptic integral of the first kind. In
optical communication such solution has been claimed~\cite{karta3,karta5}) to
represents a steady-state structure describing a train of pulses, which equal
temporal separation coincides with the period $\tau_0$ of the $dn$ function. It
is useful to end the current discussion on properties of the $dn$ function by
remarking that the one-to-one correspondance between~(\ref{a2}) and the
piecewise multi-soliton signal intensity defined in~(\ref{a2}), can easily be
established by setting $t_n=t_0 + n\tau_0$ with $t_0$ an initial position, and
constraining all pulses to have common amplitude $Q_n$ (hence common central
frequency width $\omega_0=\omega$). \\
Now turning to the probe equation~(\ref{a1b}) we remark to start that since the
nonlinearity here proceeds from the XPM coupling to the pump, it is ready to
anticipate arguing that any nonlinear mode emerging in the probe should be
induced by the pump signal. Therefore, to ensure full account of the
steady-state feature of the pump soliton as well as the temporal multiplexing at
the fiber entry, we must also express the envelope of the probe as a steady wave
i.e.:
\begin{equation}
A_2(z, t)= u(t)\,e^{i\left[k z + \phi(t)\right]}, \label{a4}
\end{equation}
where $u(t)$ is the temporal core of the probe envelope which is actually
trapped by the pump, $k$ is the wavector for the probe modulation in the pump
trap while $\phi(t)$ is a momentum-independent temporal phase shift.
Inserting~(\ref{a4}) in~(\ref{a1b}) and taking $D>0$ we obtain:      
\begin{equation}
-\frac{\partial^2 u}{\partial \tau^2} + \left[\frac{\beta}{\xi d} \kappa^2
sn^2(\tau) -  h(k) \right]u=0, \hskip 0.3truecm \phi_t= -\frac{\lambda}{d},
\label{a5}
\end{equation}
\begin{equation}
\tau= Q (t - t_0), \hskip 0.3truecm h(k)= \frac{2\beta}{\xi d} +
\frac{2\epsilon}{Q^2 d}, \hskip 0.3truecm \epsilon= k + \frac{\lambda^2}{2d}.
\label{a5b}
\end{equation}
Now setting 
\begin{equation}
n(n + 1)= \frac{2\beta}{\xi D}, \label{a5c}
\end{equation}
equation~(\ref{a5}) becomes:
\begin{equation}
u_{\tau \tau} + \left[h(k) - n(n + 1)\kappa^2 sn^2(\tau)\right] u=0. \label{a6} 
\end{equation}
When $\kappa=1$, equation~(\ref{a6}) reduces exactly to the Associated Legendre
equation studied in refs.~\cite{manas1,manas2,rand2,rand3} in the context of a
single-soliton pump interacting with an harmonic probe. In the next section, we
investigate spectral properties of the eigenvalue equation~(\ref{a6}) for values
of $\beta/\xi D$ larger than those considered so far, in particualr for orders
of the competition between the XPM and SPM effects higher than the lowest order
discussed in~\cite{dika1}. 
\section{Spectral properties of $\vert
n(=2,3),\,l\textgreater$ soliton eigenstates}  
\subsection{The $\vert 2,l\textgreater$ soliton modes}
\begin{figure*}
\begin{minipage}[c]{0.51\textwidth}
\includegraphics[width=3.0in,height=1.65in]{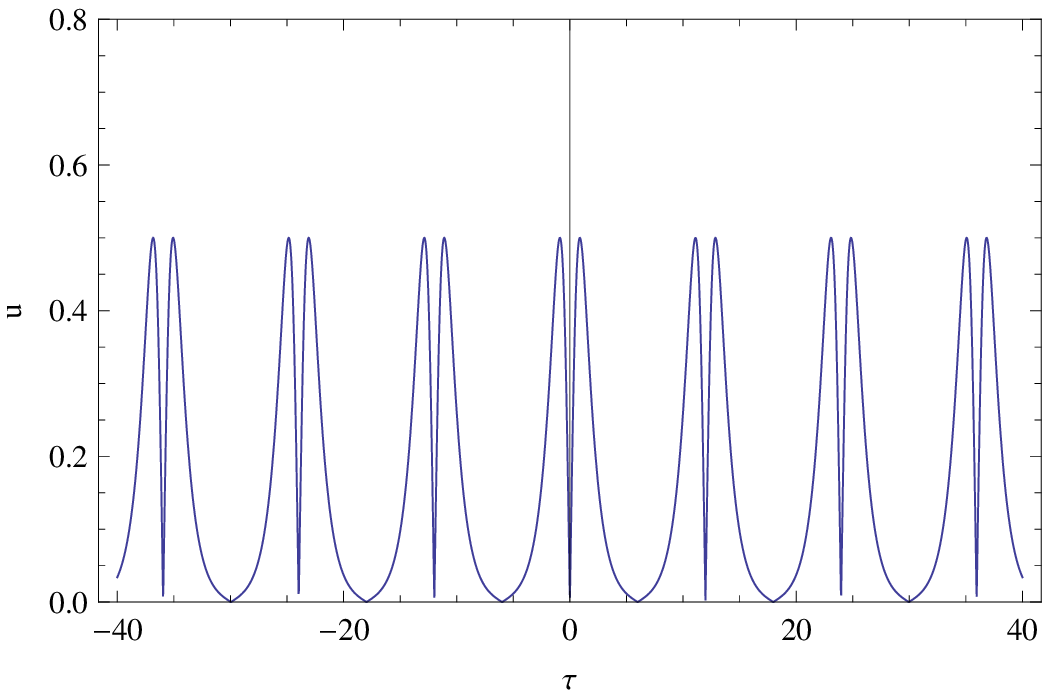}
\end{minipage}%
\begin{minipage}[c]{0.51\textwidth}
\includegraphics[width=3.0in,height=1.65in]{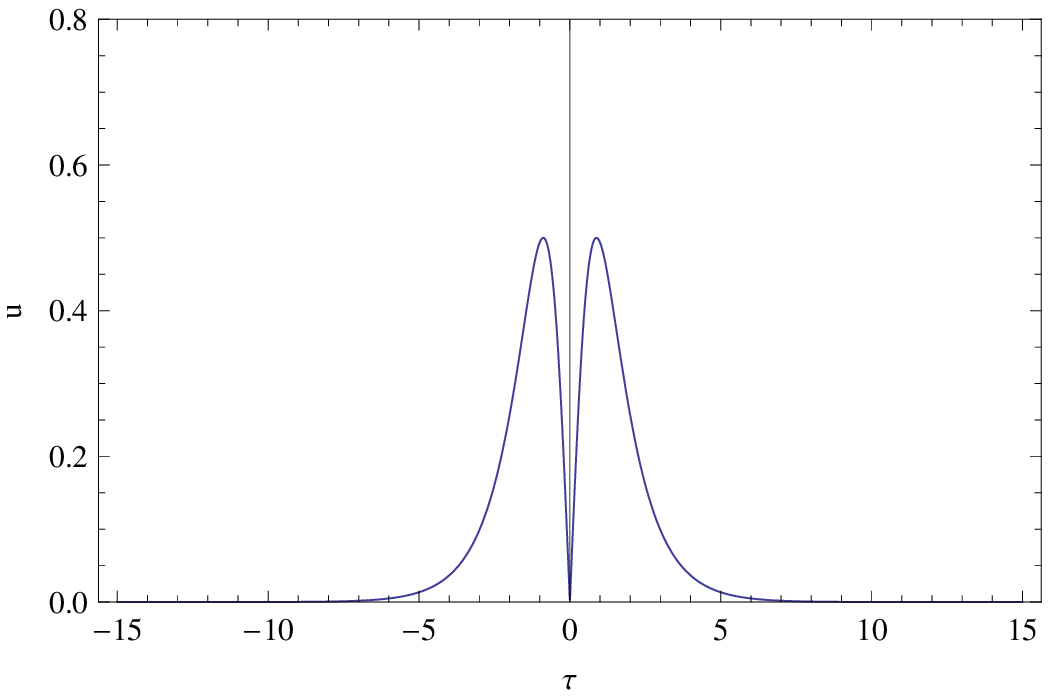}
\end{minipage} \\
\begin{minipage}[c]{0.51\textwidth}
\includegraphics[width=3.0in,height=1.65in]{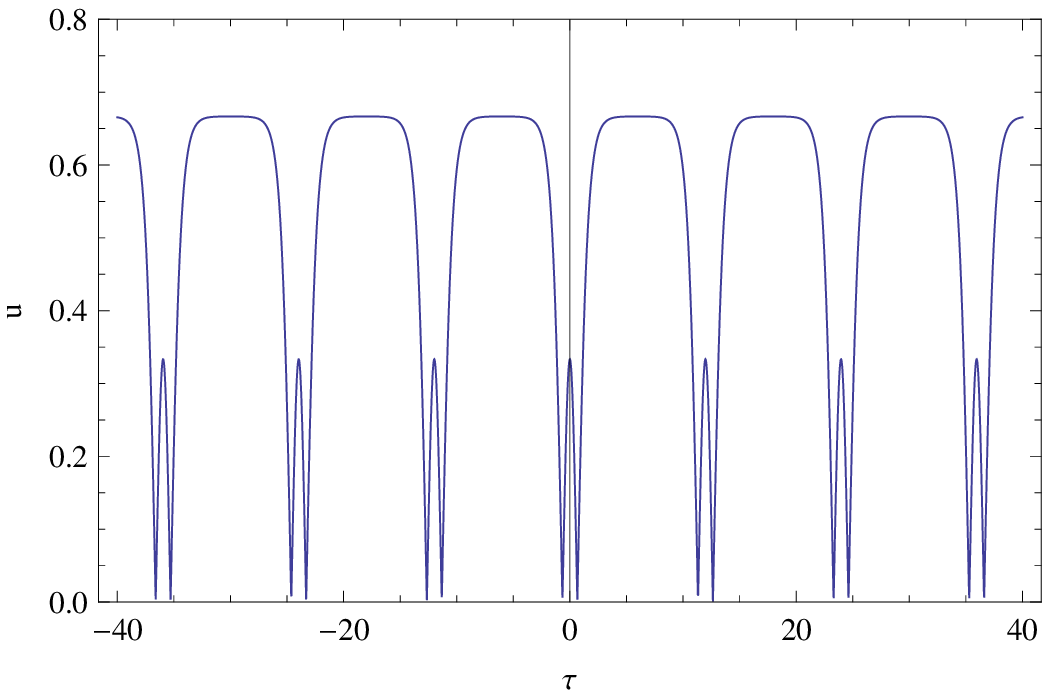}
\end{minipage}%
\begin{minipage}[c]{0.51\textwidth}
\includegraphics[width=3.0in,height=1.65in]{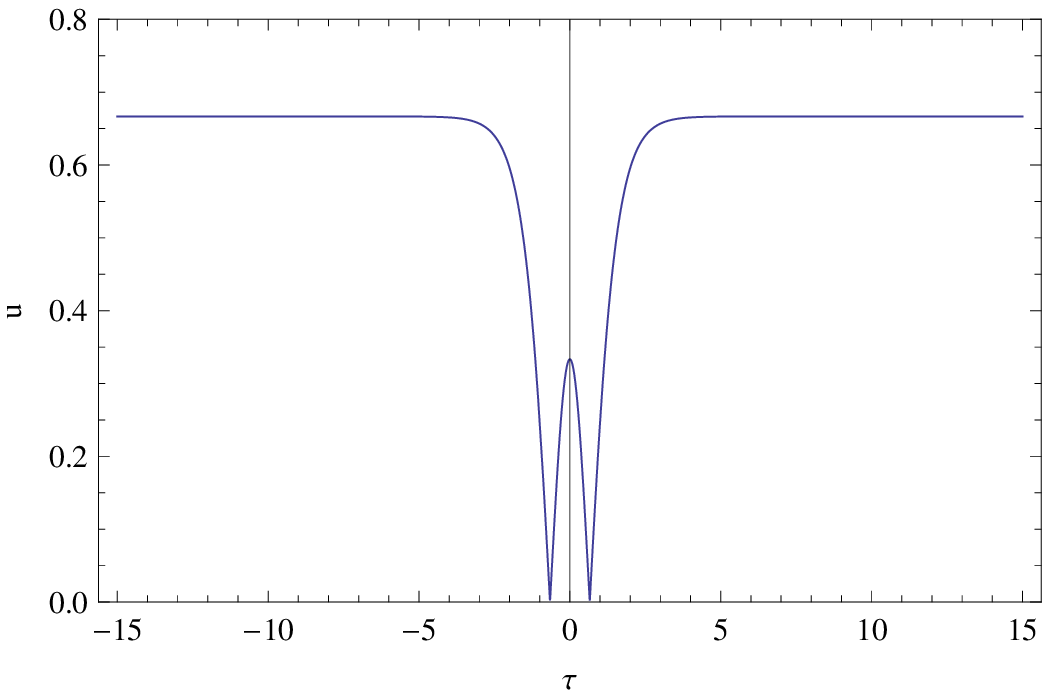}
\end{minipage} \\
\begin{minipage}[c]{0.51\textwidth}
\includegraphics[width=3.0in,height=1.65in]{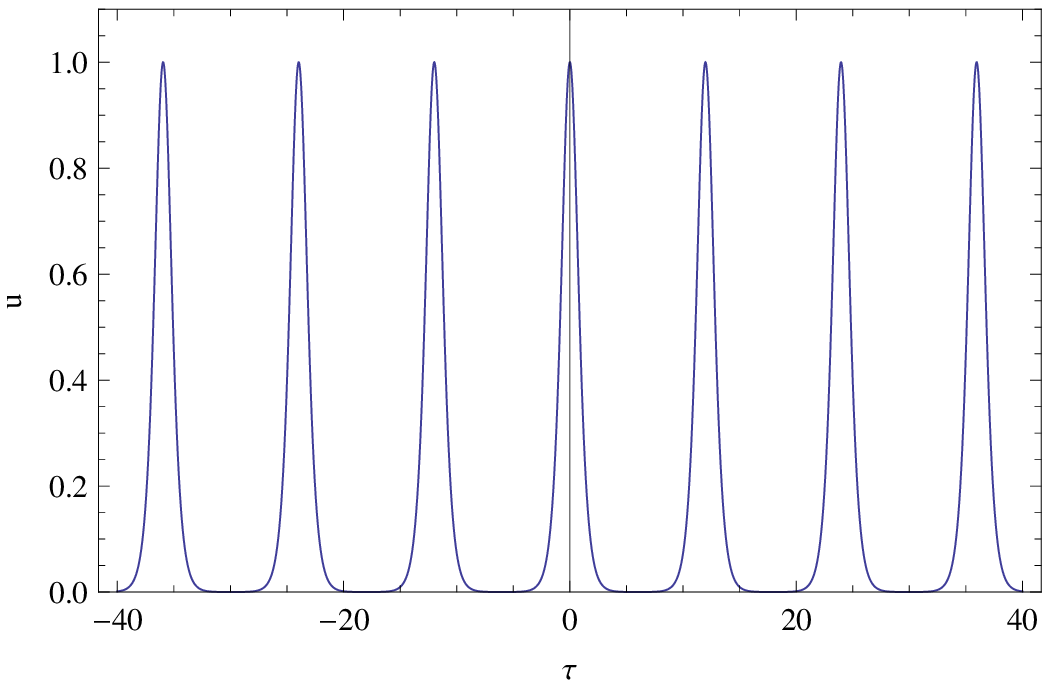}
\end{minipage}%
\begin{minipage}[c]{0.51\textwidth}
\includegraphics[width=3.0in,height=1.65in]{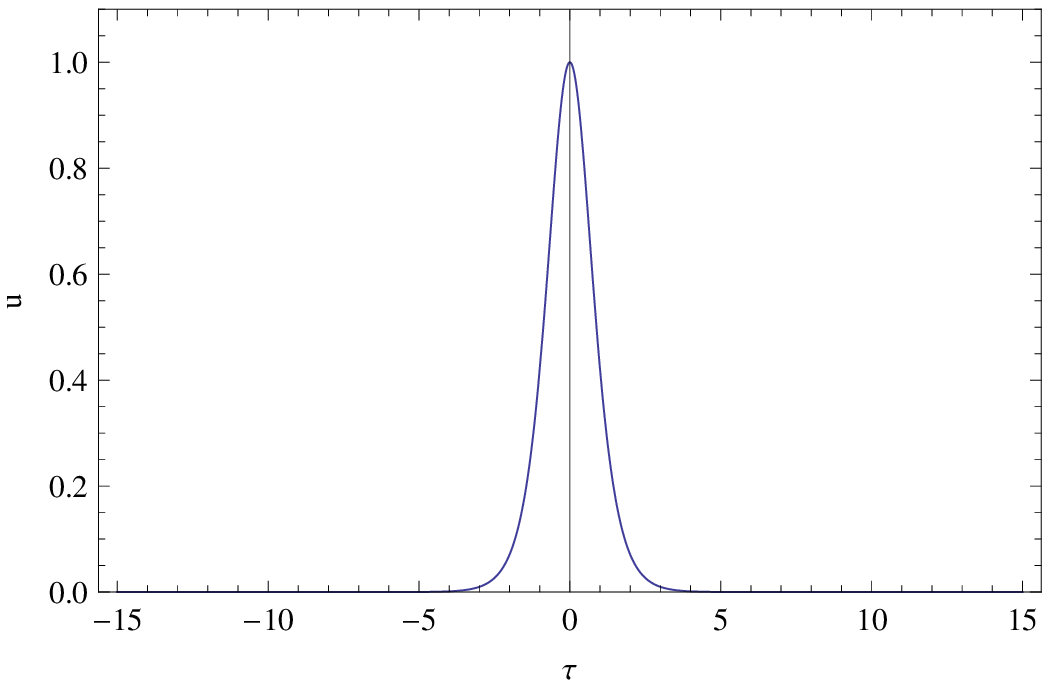}
\end{minipage} \\
\begin{minipage}[c]{0.51\textwidth}
\includegraphics[width=3.0in,height=1.65in]{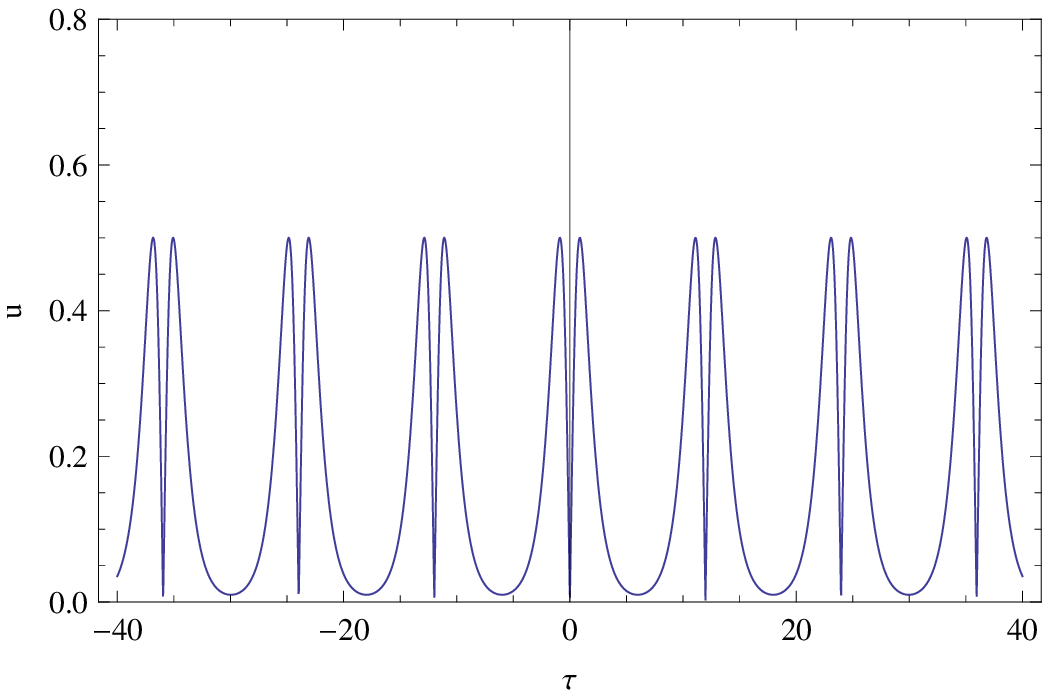}
\end{minipage}%
\begin{minipage}[c]{0.51\textwidth}
\includegraphics[width=3.0in,height=1.65in]{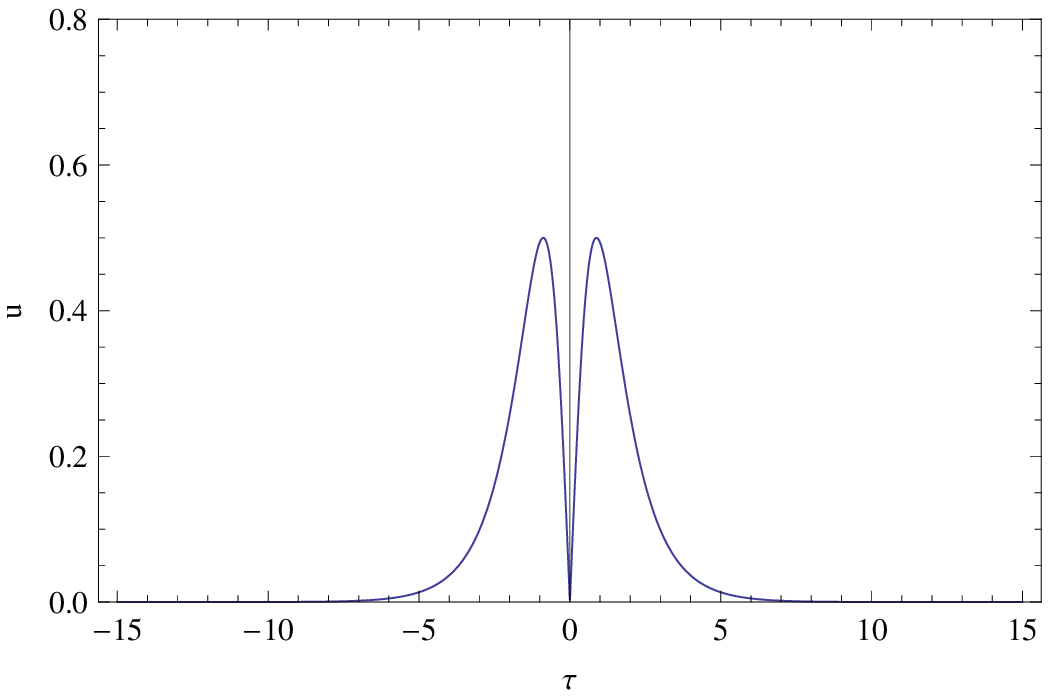}
\end{minipage} \\
\begin{minipage}[c]{0.51\textwidth}
\includegraphics[width=3.0in,height=1.65in]{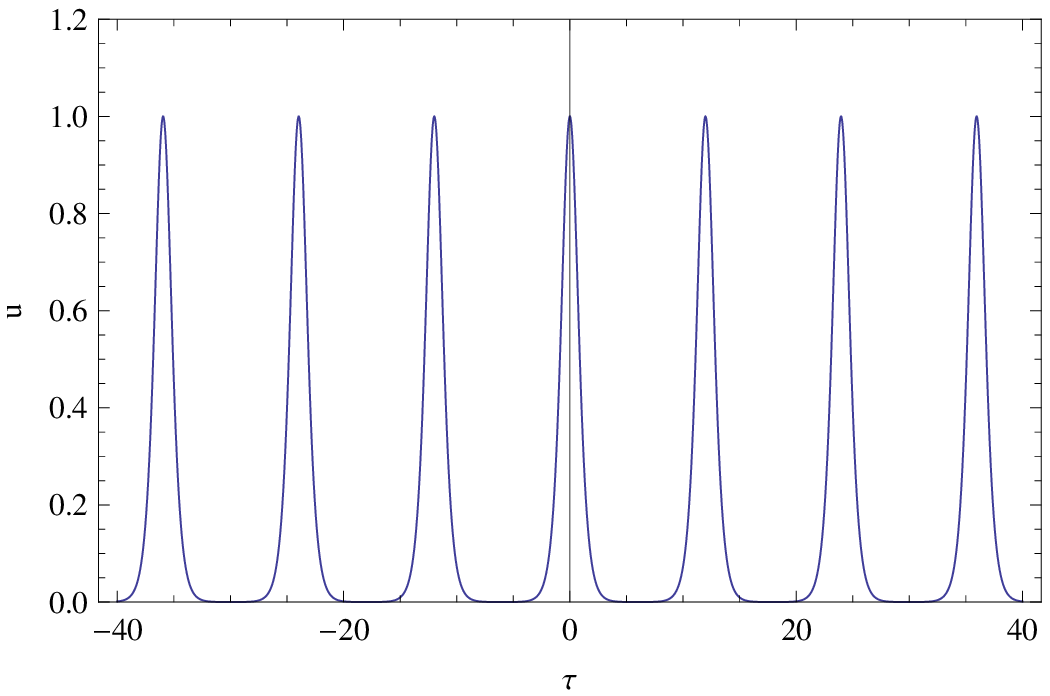}
\end{minipage}%
\begin{minipage}[c]{0.51\textwidth}
\includegraphics[width=3.0in,height=1.65in]{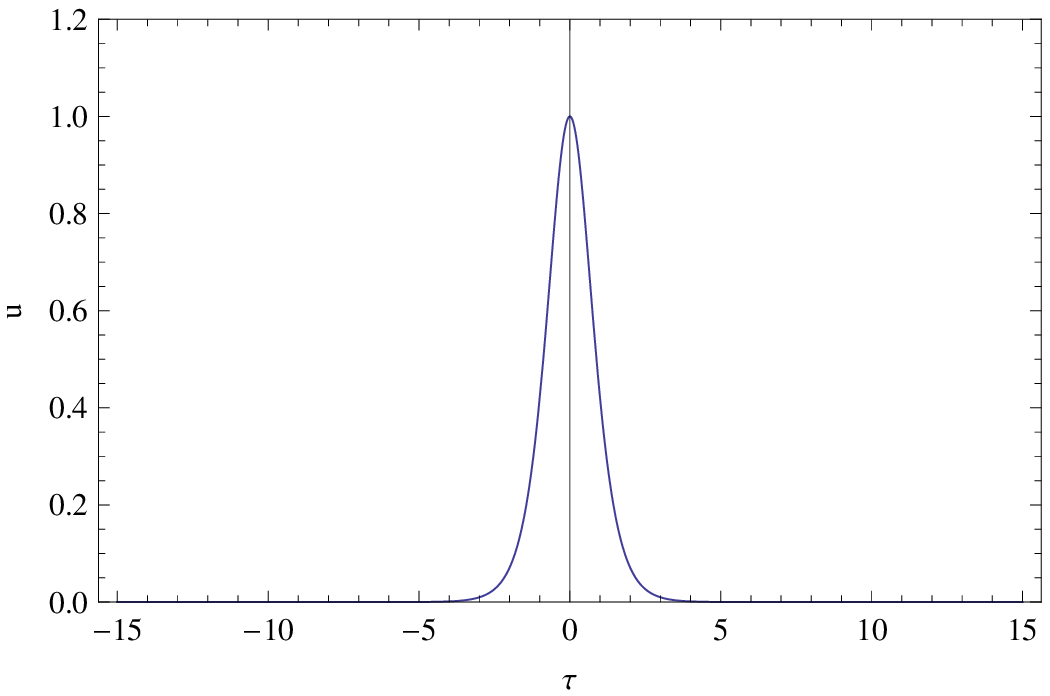}
\end{minipage} 
\caption{\label{fig:one} Intensities $\vert u(\tau)\vert$ of the five soliton
modes $\vert 2,l\textgreater$ given by~(\ref{a7a})-(\ref{a7e}), for
$\kappa=0.95$ (left graphs) and $\kappa=1$ (right graphs). Modes are displayed
from the lowest (top graph) to the highest (bottom graph) soliton eigenstates in
terms of their modulation wavectors.}
\end{figure*}

\begin{figure}
\includegraphics[width=3.0in,height=2.5in]{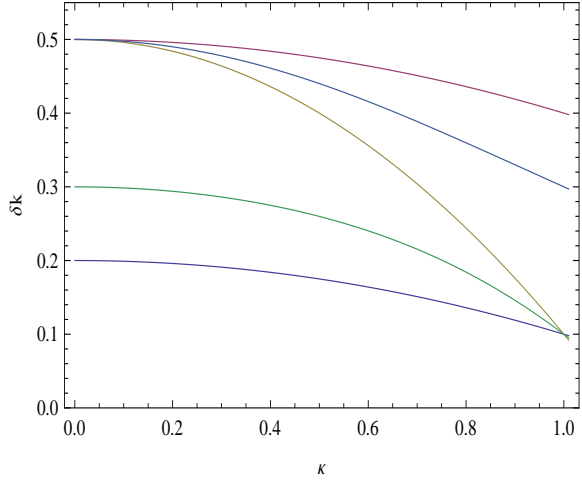}
\caption{\label{fig:two} Spectrum of the five localized soliton
eigenstates~(\ref{a7a})-(\ref{a7e}), representing their modulation wavectors as
a function of $\kappa$. The bottom curve corresponds to the lowest soliton
eigenstate and the top curve to the highest soliton eigenstate of the trapped
probe.}
\end{figure}

When $\beta=3\,\xi D$, formula~(\ref{a5c}) yields $n=2$ and the spectral problem
for the probe in the pump trap takes the explicit form:
\begin{equation}
u_{\tau \tau} + \left[h(k) - 6\kappa^2 sn^2(\tau)\right] u=0. \label{a6a} 
\end{equation}
Equation~(\ref{a6a}), which is member of the so-called Lam\'e eigenvalue
problem~\cite{dika2,ars}, is known to admit both extended-wave and
localized-wave solutions. For the problem under consideration localized-wave
solutions are best appropriate for they represent probe structures that involve
low-momentum exchanges with the pump signal. Moreover, their localized shapes
are key asset in our quest for stable long-standing signals with soliton
features. In the specific case of equation~(\ref{a6a}) there are exactly five
distinct localized modes that are induced by the soliton trap in the probe
groundstate. These five modes, which can be induced simultaneously in the probe
for a pump of sufficiently high intensity, form a complete set and the associate
spectrum consists of the following eigenstates listed from the lowest to the
highest modes (in terms of the magntitudes of their respective modulation
wavectors $k_l$)~\cite{dika2,ars}:  \\
$\vert 2,1\textgreater$ soliton mode:
\begin{eqnarray}
u_1(\tau)&=& u_{1,0}(\kappa)\, cn\left(\tau - \tau_0\right)\,sn\left(\tau -
\tau_0\right), \nonumber \\ 
k_1(\kappa)&=& k_{\lambda} + Q^2D\,(2-\kappa^2)/2, \label{a7a} 
\end{eqnarray}
$\vert 2,2\textgreater$ soliton mode: 
\begin{eqnarray}
u_2(\tau)&=& u_{2,0}(\kappa)\, \left[sn^2\left(\tau - \tau_0\right) -
\frac{1+\kappa^2 -\sqrt{1-\kappa^2(1-\kappa^2)}}{3\kappa^2}\right], \nonumber
\\ 
k_2(\kappa)&=& k_{\lambda} + \frac{Q^2D}{2}\,\left[2(2-\kappa^2) -
\sqrt{1-\kappa^2(1-\kappa^2)}\right], \label{a7c} 
\end{eqnarray}
$\vert 2,3\textgreater$ soliton mode: 
\begin{eqnarray}
u_3(\tau)&=& u_{3,0}(\kappa)\, \left[sn^2\left(\tau - \tau_0\right) -
\frac{1+\kappa^2 + \sqrt{1-\kappa^2(1-\kappa^2)}}{3\kappa^2}\right], \nonumber
\\ 
k_3(\kappa)&=& k_{\lambda} + \frac{Q^2D}{2}\,\left[2(2-\kappa^2) +
\sqrt{1-\kappa^2(1-\kappa^2)}\right], \label{a7d} 
\end{eqnarray}
$\vert 2,4\textgreater$ soliton mode: 
\begin{eqnarray}
u_4(\tau)&=& u_{4,0}(\kappa)\, dn\left(\tau - \tau_0\right)\,sn\left(\tau -
\tau_0\right), \nonumber \\ 
k_4(\kappa)&=& k_{\lambda} + Q^2D\,(5-4\kappa^2)/2, \label{a7b} 
\end{eqnarray}
$\vert 2,5\textgreater$ soliton mode: 
\begin{eqnarray}
u_5(\tau)&=& u_{5,0}(\kappa)\, cn\left(\tau - \tau_0\right)\,dn\left(\tau -
\tau_0\right), \nonumber \\ 
k_5(\kappa)&=& k_{\lambda} + Q^2D\,(5-\kappa^2)/2. \label{a7e} 
\end{eqnarray}
In the above formula $cn$ and $sn$ are Jacobi elliptic functions while
$k_{\lambda}=\lambda^2/2D$ is a wavector shift associate with the temporal
walk-off $\lambda$. According to expressions of the characteristic wavectors of
the five soliton modes, this wavector shift is uniform and constant so that the
temporal walk-off is not relevant to the generation mechanism of the localized
modes, but strictly the competition between the XPM and SPM couplings.  \\
In figure~\ref{fig:one}, we sketched profiles of the intensities $\vert
u(\tau)\vert$ of the five soliton modes obtained above for arbitrary values of
their normalization amplitudes. The left graphs are intensity profiles of the
induced-soliton signals for $\kappa=0.95$, while the right graphs represent same
quantities for $\kappa=1$. 
According to the left graphs of figure~\ref{fig:one}, the five localized modes
have periodic-wavetrain features similar to the pump structure when $0\leq
\kappa <1$. Moreover, as in this last structure where the fundamental signal is
pulse shaped, the five modes are pulse-core periodic signals though with
distinct periodicities. Indeed, the first (i.e. lowest) and fourth modes are
period-one double-pulse soliton trains, the second is a period-two one-pulse
soliton train whereas the third and fifth modes are period-one one-pulse soliton
trains. Shapes of the fundamental solitons characterizing the five soliton modes
are displayed in the right graphs of figure~\ref{fig:one}, they are consistent
both with the general analytical expressions~(\ref{a7a})-(\ref{a7e}) when
plotted for $\kappa= 1$, and with the following explicit formula obtained from
these solutions in the limit $\kappa\rightarrow 1$: \\
\begin{eqnarray}
u_1(\tau)&\propto & sech\left(\tau - \tau_0\right)\,tanh\left(\tau -
\tau_0\right), \nonumber \\ 
k_1&=& k_{\lambda} + Q^2D/2, \label{a8a} 
\end{eqnarray}
\begin{eqnarray}
u_2(\tau)&\propto & \left[\frac{2}{3} - sech^2\left(\tau - \tau_0\right)\right],
\nonumber \\ 
k_2&=& k_{\lambda} + Q^2D/2, \label{a8c} 
\end{eqnarray}
\begin{eqnarray}
u_3(\tau)&\propto & - sech^2\left(\tau - \tau_0\right), \nonumber \\ 
k_3&=& k_{\lambda} + 3Q^2D/2, \label{a8d} 
\end{eqnarray}
\begin{eqnarray}
u_4(\tau)&\propto& sech\left(\tau - \tau_0\right)\,tanh\left(\tau -
\tau_0\right), \nonumber \\ 
k_4&=& k_{\lambda} + Q^2D/2, \label{a8b} 
\end{eqnarray}
\begin{eqnarray}
u_5(\tau)&\propto & sech^2\left(\tau - \tau_0\right), \nonumber \\ 
k_5&=& k_{\lambda} + 2Q^2D. \label{a8e} 
\end{eqnarray}
To give a complete description of spectral properties of the $\vert
2,l\textgreater$ probe soliton modes,  we also plotted the variations of their
modulation wavectors with the paramater $\kappa$. In effect, the modulus
$\kappa$ of Jacobi elliptic functions emerged above as governing the stability
of wavetrain structures of both the pump and the induced probe solitons, as well
as their decay toward characteristic fundamental soliton modes. Since modes are
characterized by their modulation wavectors within the pump trap and given that
the modulation wavector determines the amount of energy cost to the pump for
mode formation and stable, we expect the variations of the different
$k_{l=1,...,5}$ obtained above to provide relevant additional insight onto
spectral properties of the five modes. \\
Figure~(\ref{fig:two}) displays the wavector spectrum incompassing the five
probe modes~(\ref{a7a})-(\ref{a7e}), where wavectors are taken relative to the
common walk-off-induced uniform shift $k_{\lambda}=\lambda^2/2D$ of the whole
spectrum as reflected by analytical expressions of the quantitites
$k_{l=1,...,5}$ derived above.
Two most stricking features emerging from the graph, that traduce relevant
physical implications of the variation of wavectors with $\kappa$ on the mode
stability, are the decrease of their modulation wavectors when they decay from
the wavetrain structure to their respective single fundamental-soliton states,
and their degenerate features in the two limit values of $\kappa$. Recall that
when $\kappa$ tends to zero, the Jacobi elliptic functions reduce to harmonic
waves which are intrinsically linear.
\subsection{The $\vert 3,l\textgreater$ soliton modes}
\begin{figure*}
\begin{minipage}[c]{0.51\textwidth}
\includegraphics[width=3.0in,height=1.9in]{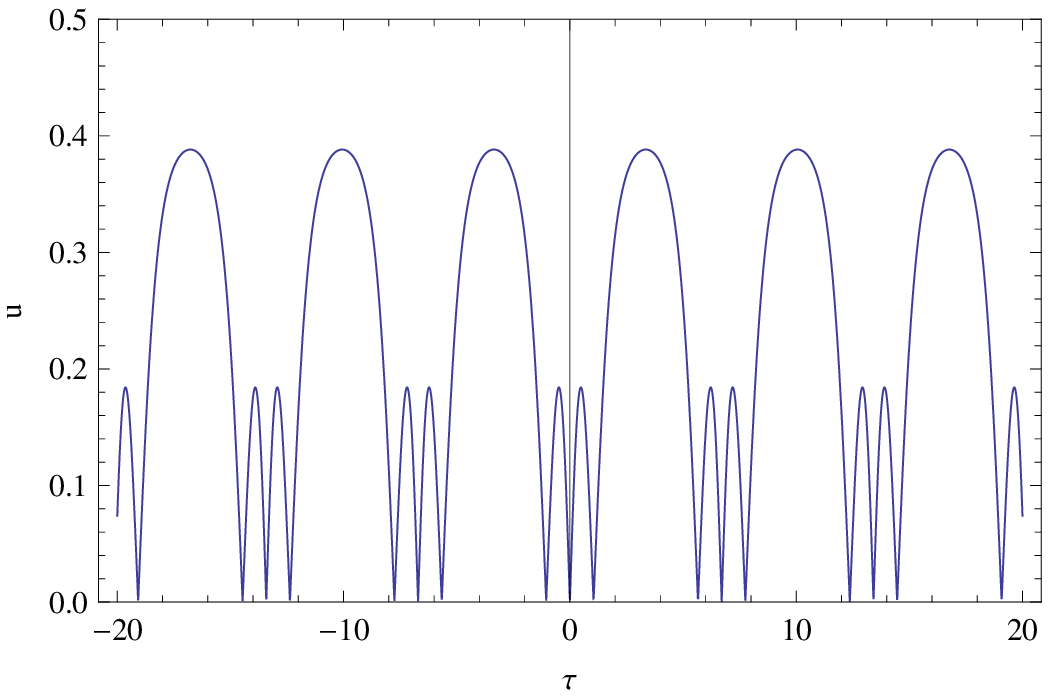}
\end{minipage}%
\begin{minipage}[c]{0.51\textwidth}
\includegraphics[width=3.0in,height=1.9in]{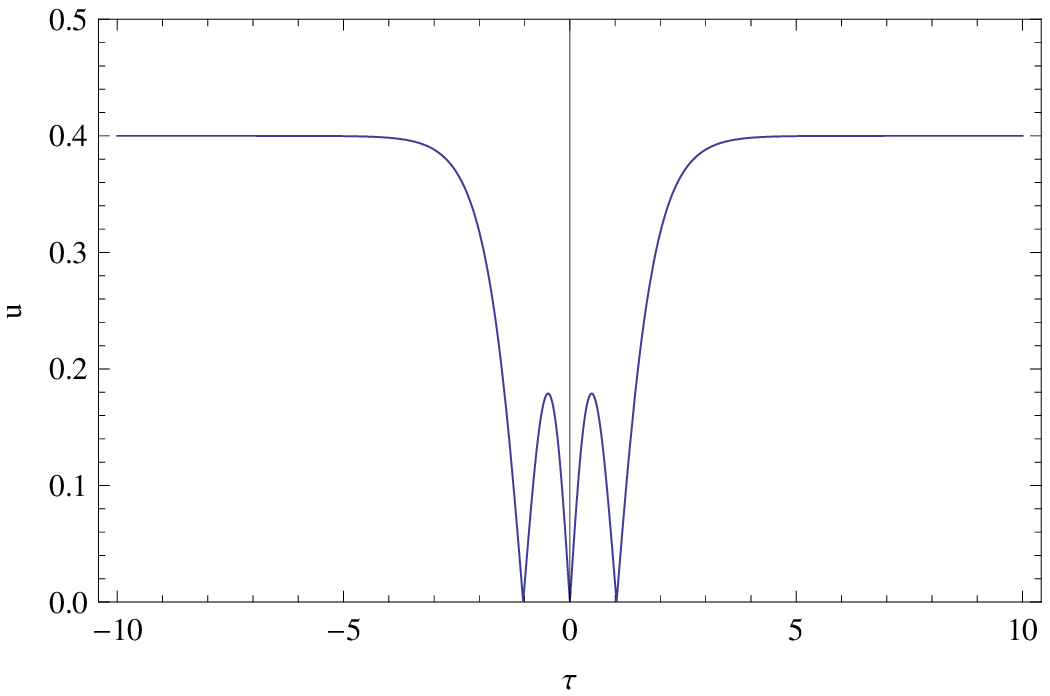}
\end{minipage} \\
\begin{minipage}[c]{0.51\textwidth}
\includegraphics[width=3.0in,height=1.9in]{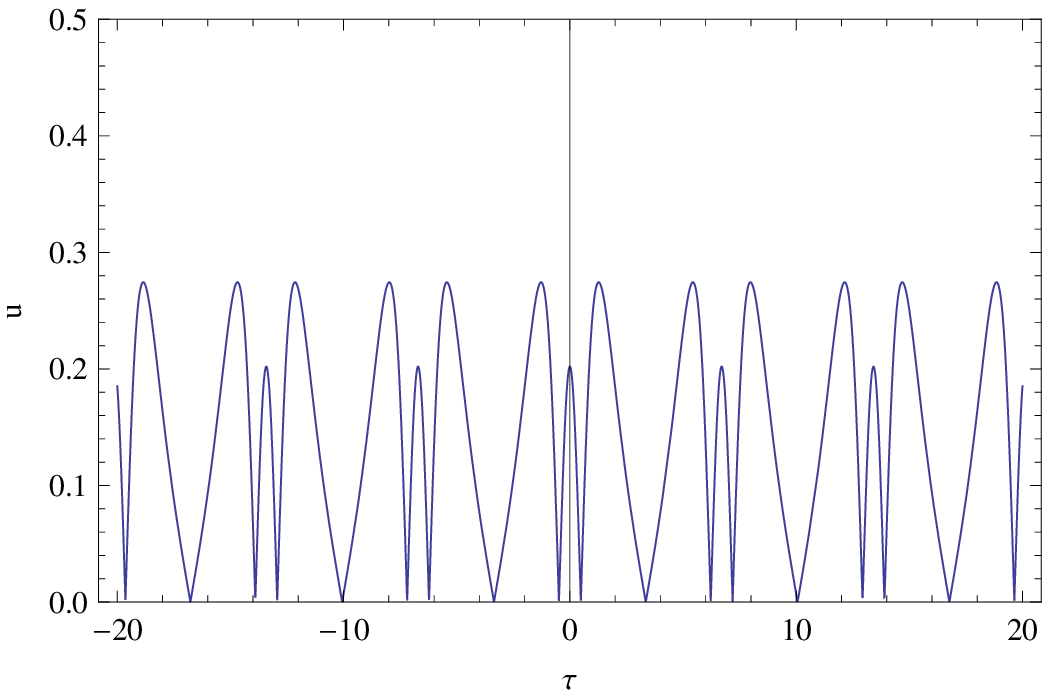}
\end{minipage}%
\begin{minipage}[c]{0.51\textwidth}
\includegraphics[width=3.0in,height=1.9in]{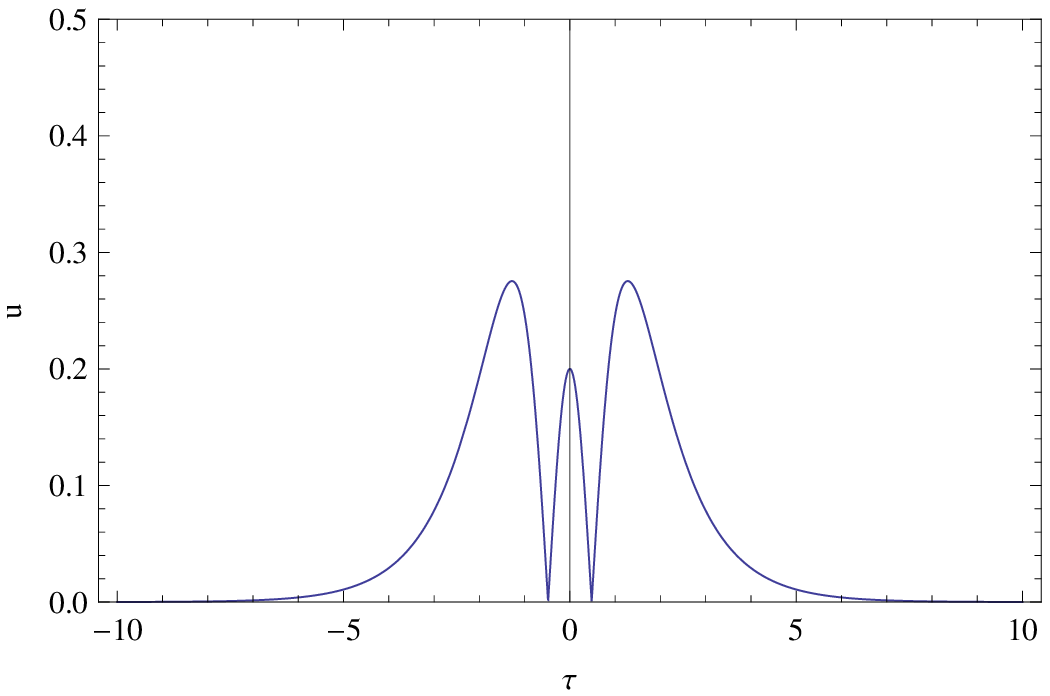}
\end{minipage} \\
\begin{minipage}[c]{0.51\textwidth}
\includegraphics[width=3.0in,height=1.9in]{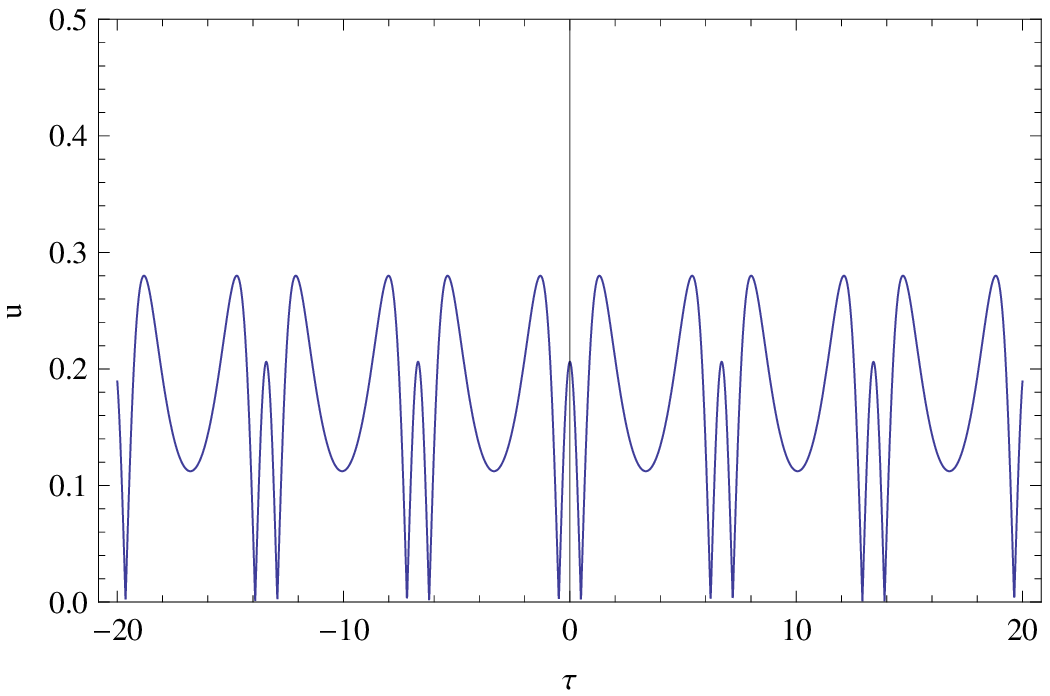}
\end{minipage}%
\begin{minipage}[c]{0.51\textwidth}
\includegraphics[width=3.0in,height=1.9in]{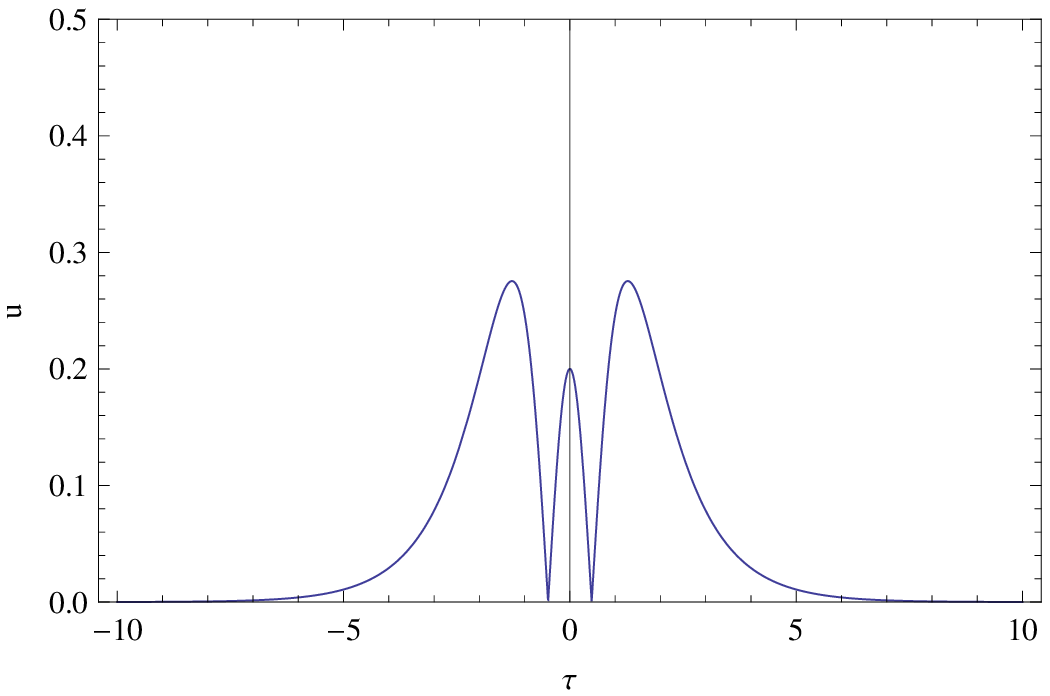}
\end{minipage}
\caption{\label{fig:three} Intensities $\vert u(\tau)\vert$ of the soliton modes
$\vert 3,l\textgreater$ given by~(\ref{a10a})-(\ref{a10b}), for $\kappa=0.95$
(left graphs) and $\kappa=1$ (right graphs). Modes are displayed from the lowest
(top graph) to the highest (bottom graph) soliton eigenstates in terms of their
modulation wavectors.}
\end{figure*}

\begin{figure*}\centering
\begin{minipage}[c]{0.51\textwidth}
\includegraphics[width=3.0in,height=1.9in]{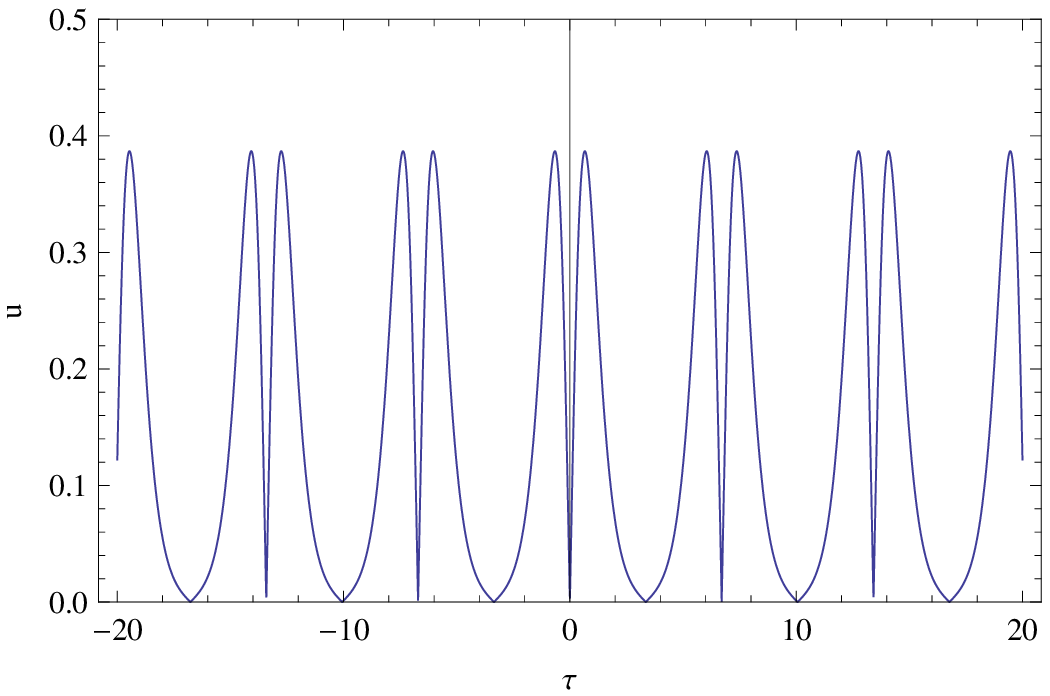}
\end{minipage}%
\begin{minipage}[c]{0.51\textwidth}
\includegraphics[width=3.0in,height=1.9in]{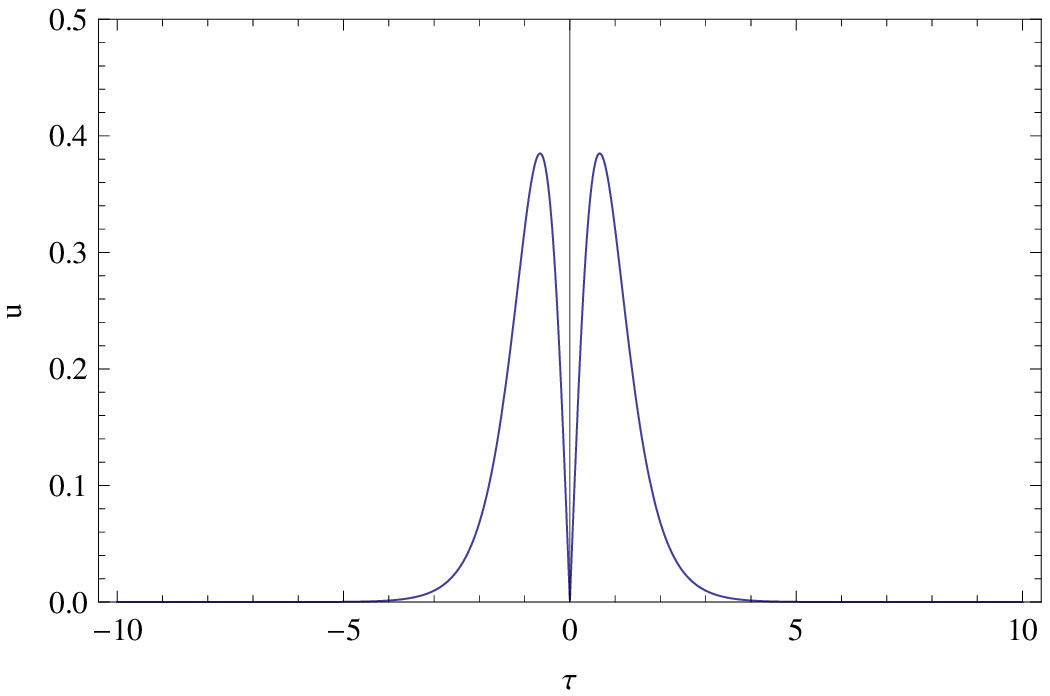}
\end{minipage}\\
\begin{minipage}[c]{0.51\textwidth}
\includegraphics[width=3.0in,height=1.9in]{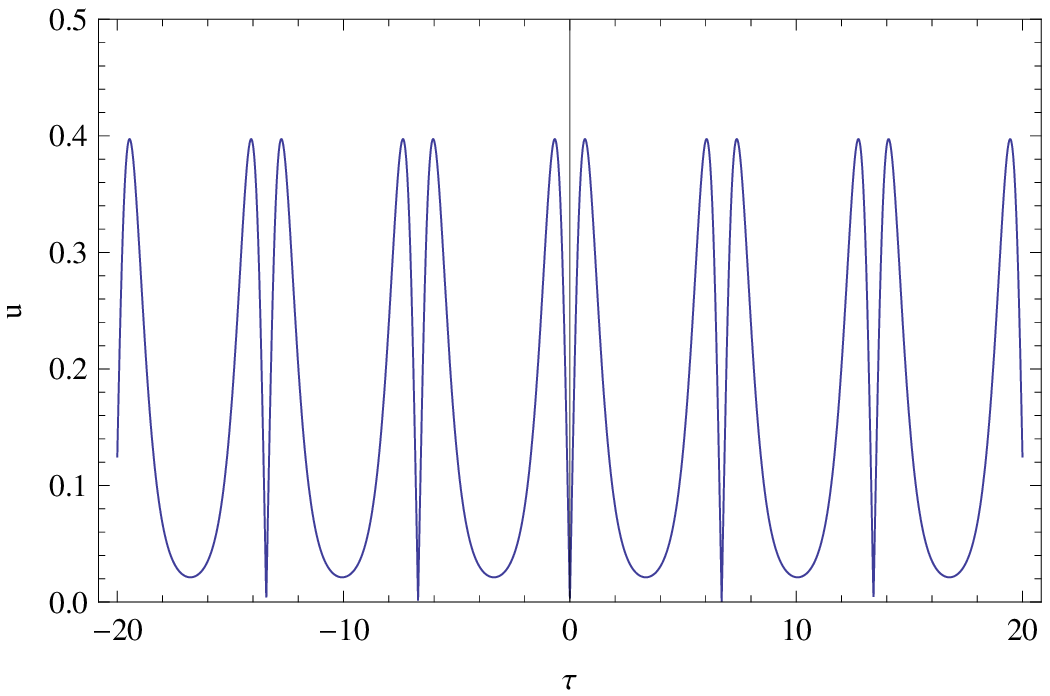}
\end{minipage}%
\begin{minipage}[c]{0.51\textwidth}
\includegraphics[width=3.0in,height=1.9in]{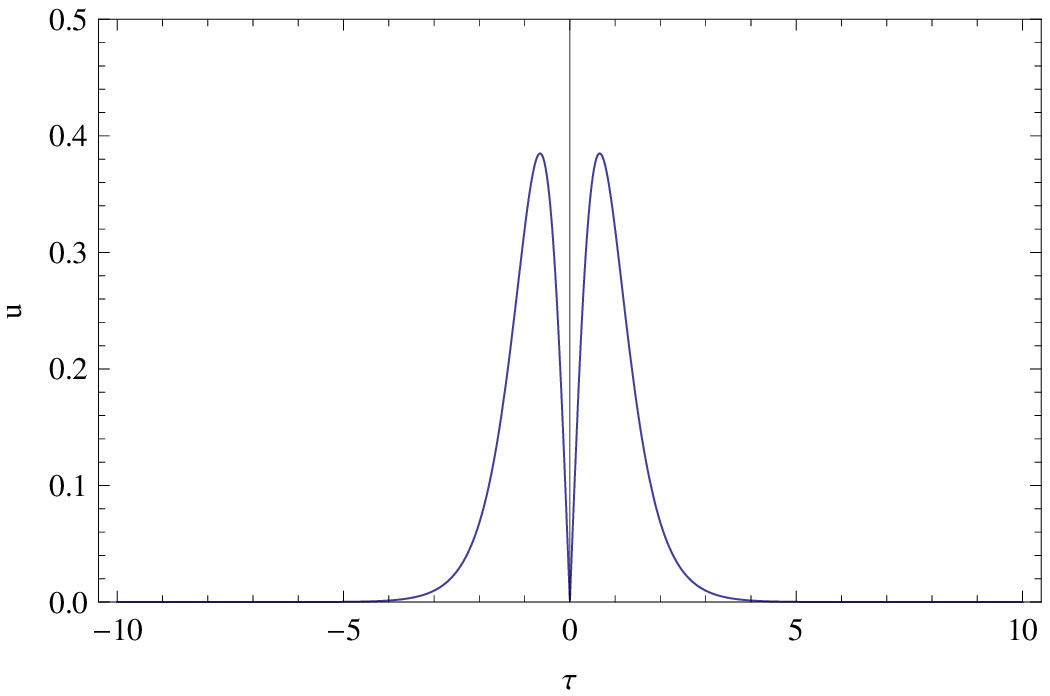}
\end{minipage}\\
\begin{minipage}[c]{0.51\textwidth}
\includegraphics[width=3.0in,height=1.9in]{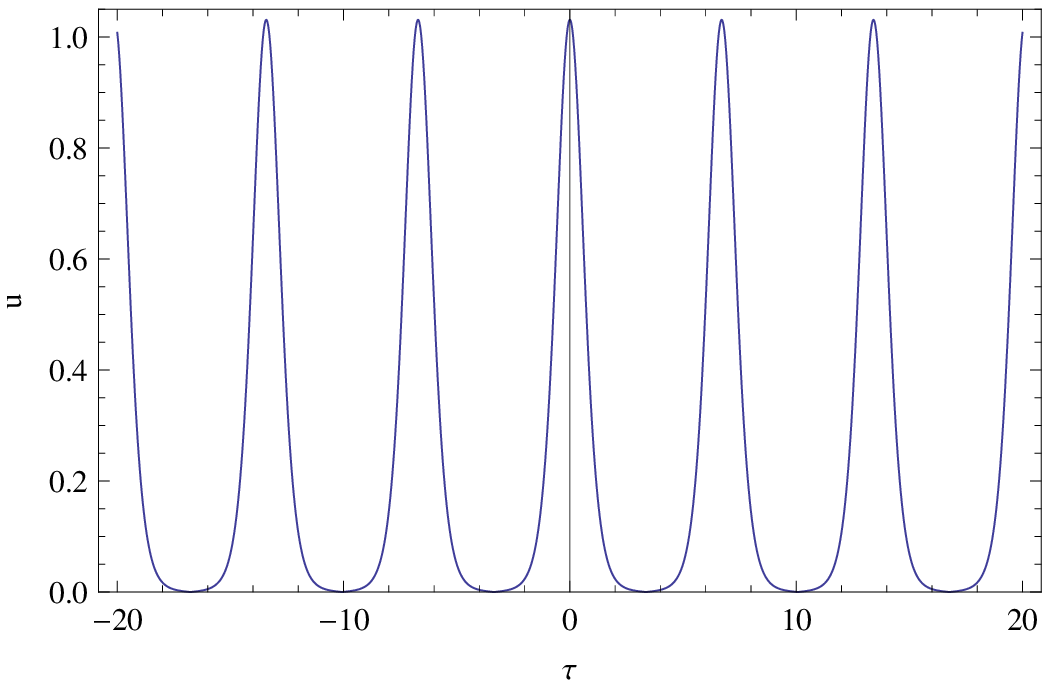}
\end{minipage}%
\begin{minipage}[c]{0.51\textwidth}
\includegraphics[width=3.0in,height=1.9in]{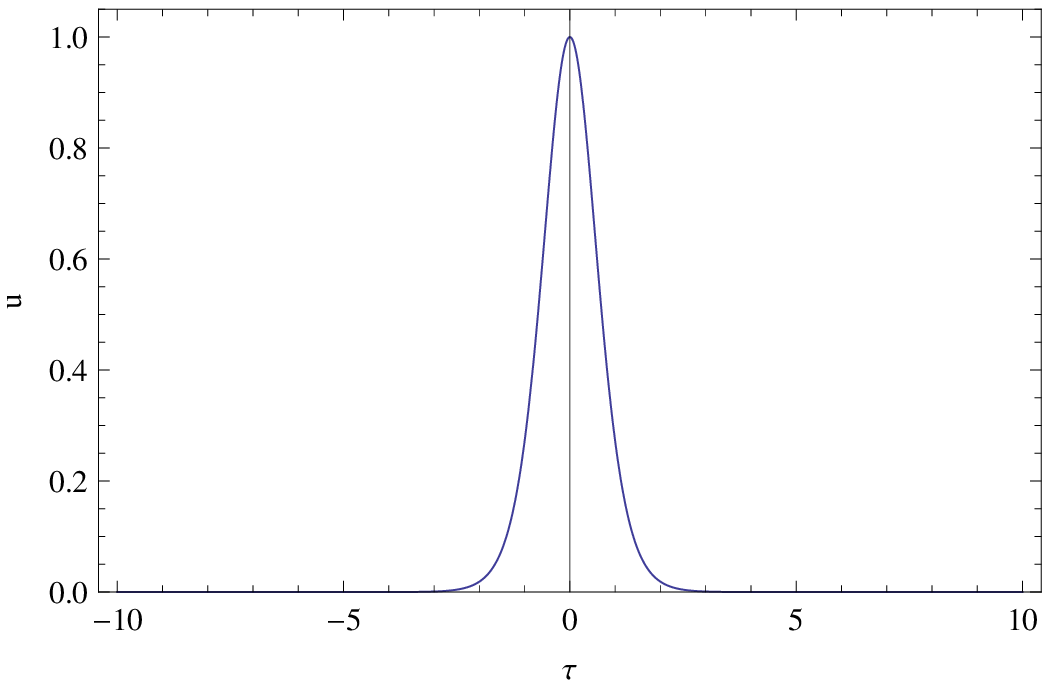}
\end{minipage} \\
\begin{minipage}[c]{0.51\textwidth}
\includegraphics[width=3.0in,height=1.9in]{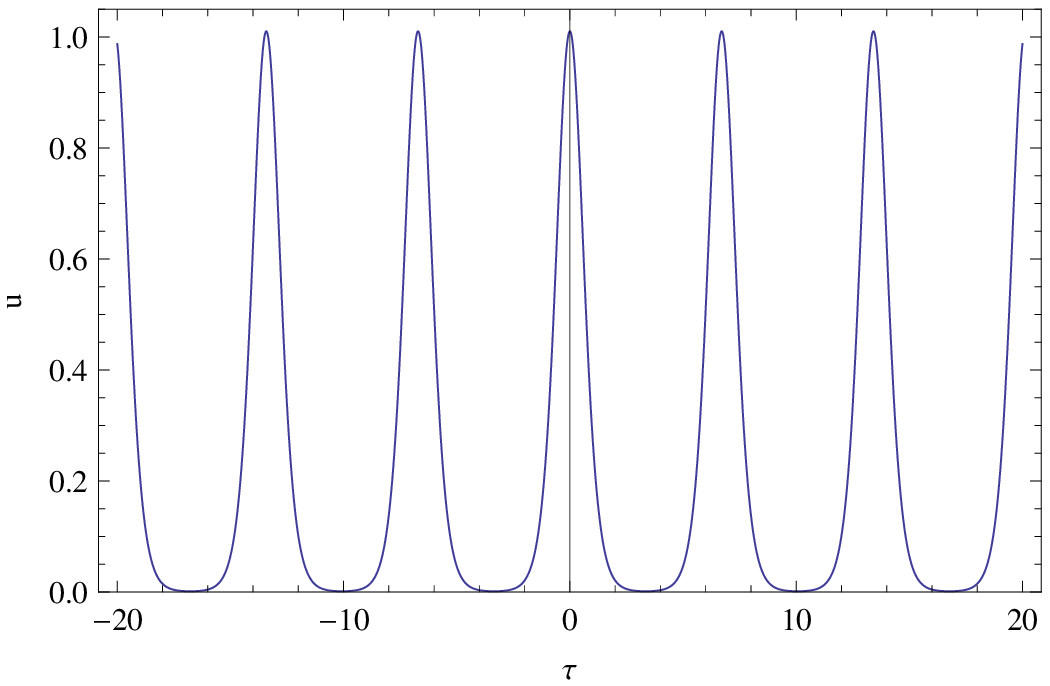}
\end{minipage}%
\begin{minipage}[c]{0.51\textwidth}
\includegraphics[width=3.0in,height=1.9in]{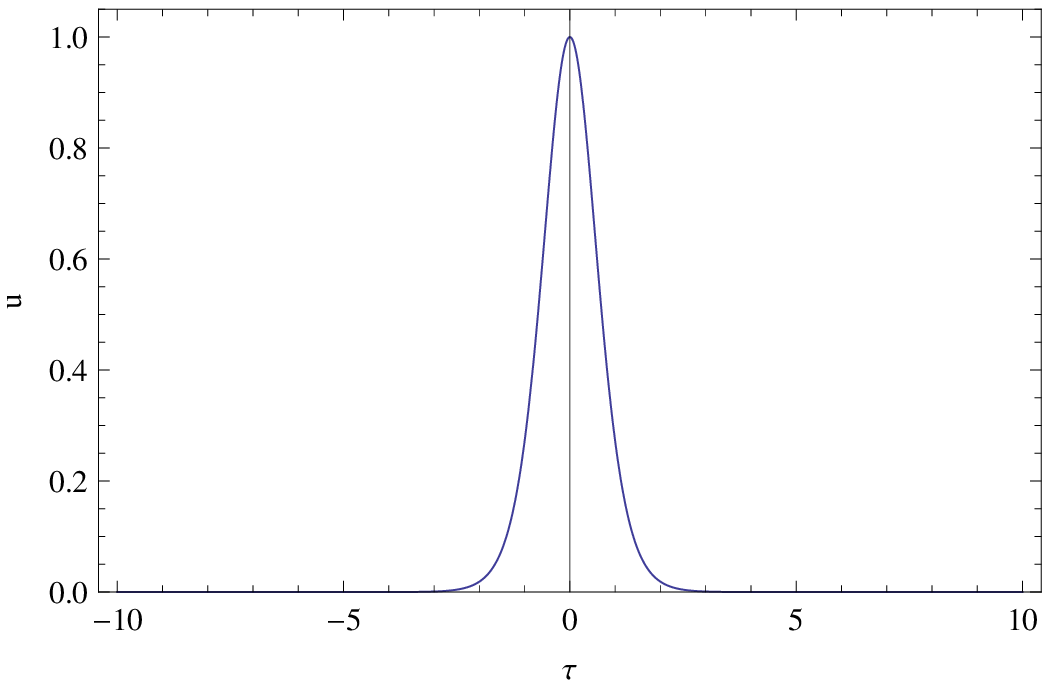}
\end{minipage}
\caption{\label{fig:four} Intensities $\vert u(\tau)\vert$ of the soliton modes
$\vert 3,l\textgreater$ given by~(\ref{a10e})-(\ref{a10f}), for $\kappa=0.95$
(left graphs) and $\kappa=1$ (right graphs). Modes are displayed from the lowest
(top graph) to the highest (bottom graph) soliton eigenstates in terms of their
modulation wavectors.}
\end{figure*}

\begin{figure}
\includegraphics[width=3.0in,height=2.5in]{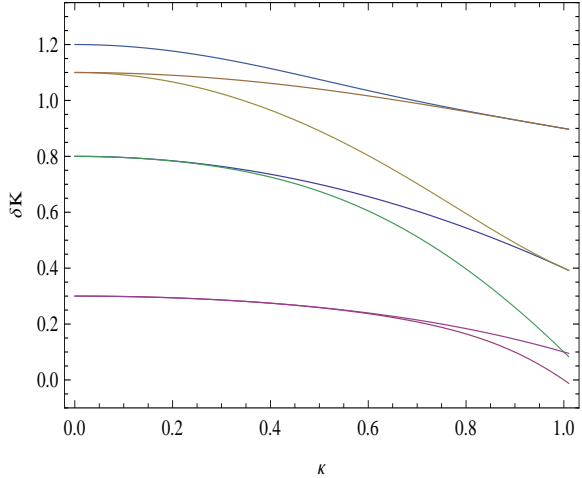}
\caption{\label{fig:five} Spectrum of the seven localized soliton
eigenstates~(\ref{a10a})-(\ref{a10f}), representing their modulation wavectors
as a function of $\kappa$. The bottom curve corresponds to the lowest soliton
eigenstate and the top curve to the highest soliton eigenstate of the trapped
probe.}
\end{figure}

When $n=3$ the ratio of the XPM to the SPM coupling strengths takes the value
$\beta=6\xi D$, corresonding to a stronger XPM effect exerted by the pump on the
probe. The resulting spectral problem for the trapped probe now reads:
\begin{equation}
u_{\tau \tau} + \left[h(k) - 12\kappa^2 sn^2(\tau)\right] u=0. \label{a9a} 
\end{equation}
For this spectral problem, the localized-mode spectrum bursts into seven
different eigenstates given by~\cite{ars}: \\  
$\vert 3,1\textgreater$ soliton mode:
\begin{eqnarray}
u_1(\tau)&=& u_{1,0}(\kappa)\, \left[sn^2\left(\tau - \tau_0\right) -
\frac{2(1+\kappa^2)
-\sqrt{4-7\kappa^2+4\kappa^4}}{5\kappa^2}\right]\,sn\left(\tau - \tau_0\right),
\nonumber \\ 
k_1(\kappa)&=& k_{\lambda} + \frac{Q^2D}{2}\left[7 - 5\kappa^2 -
2\sqrt{4-7\kappa^2+4\kappa^4}\right], \label{a10a} 
\end{eqnarray}
$\vert 3,2\textgreater$ soliton mode:
\begin{eqnarray}
u_2(\tau)&=& u_{2,0}(\kappa)\, \left[sn^2\left(\tau - \tau_0\right) - \frac{2
+\kappa^2 -\sqrt{4-\kappa^2(1-\kappa^2)}}{5\kappa^2}\right]\,cn\left(\tau -
\tau_0\right), \nonumber \\ 
k_2(\kappa)&=& k_{\lambda} + \frac{Q^2D}{2}\left[7 - 2\kappa^2 -
2\sqrt{4-\kappa^2(1-\kappa^2)}\right], \label{a10c} 
\end{eqnarray}
$\vert 3,3\textgreater$ soliton mode:
\begin{eqnarray}
u_3(\tau)&=& u_{3,0} \left[sn^2\left(\tau - \tau_0\right) - \frac{1+2\kappa^2
-\sqrt{1-\kappa^2+4\kappa^4}}{5\kappa^2}\right]\,dn\left(\tau - \tau_0\right),
\nonumber \\ 
k_3(\kappa)&=& k_{\lambda} + \frac{Q^2D}{2}\left[10 - 5\kappa^2 - 2\sqrt{1-\kappa^2
+4\kappa^4}\right], \label{a10b} 
\end{eqnarray}
$\vert 3,4\textgreater$ soliton mode:
\begin{eqnarray}
u_4(\tau)&=& u_{4,0}(\kappa)\, cn\left(\tau - \tau_0\right)\,dn\left(\tau -
\tau_0\right)\,sn\left(\tau - \tau_0\right), \nonumber \\ 
k_4(\kappa)&=& k_{\lambda} + 2Q^2D(2-\kappa^2), \label{a10e} 
\end{eqnarray}
$\vert 3,5\textgreater$ soliton mode:
\begin{eqnarray}
u_5(\tau)&=& u_{5,0}(\kappa)\, \left[sn^2\left(\tau - \tau_0\right) -
\frac{2(1+\kappa^2)
+\sqrt{4-7\kappa^2+4\kappa^4}}{5\kappa^2}\right]\,sn\left(\tau - \tau_0\right),
\nonumber \\ 
k_5(\kappa)&=& k_{\lambda} + \frac{Q^2D}{2}\left[7 - 5\kappa^2 +
2\sqrt{4-7\kappa^2+4\kappa^4}\right], \label{a10d} 
\end{eqnarray}
$\vert 3,6\textgreater$ soliton mode:
\begin{eqnarray}
u_6(\tau)&=& u_{6,0}(\kappa)\, \left[sn^2\left(\tau - \tau_0\right) - \frac{2
+\kappa^2 +\sqrt{4-\kappa^2(1-\kappa^2)}}{5\kappa^2}\right]\,cn\left(\tau -
\tau_0\right), \nonumber \\ 
k_6(\kappa)&=& k_{\lambda} + \frac{Q^2D}{2}\left[7 - 2\kappa^2 +
2\sqrt{4-\kappa^2(1-\kappa^2)}\right]. \label{a10g} 
\end{eqnarray}
$\vert 3,7\textgreater$ soliton mode:
\begin{eqnarray}
u_7(\tau)&=& u_{7,0}(\kappa)\, \left[sn^2\left(\tau - \tau_0\right) -
\frac{1+2\kappa^2 + \sqrt{1-\kappa^2+4\kappa^4}}{5\kappa^2}\right]\,dn\left(\tau
- \tau_0\right), \nonumber \\ 
k_7(\kappa)&=& k_{\lambda} + \frac{Q^2D}{2}\left[10 - 5\kappa^2 + 2\sqrt{1-\kappa^2
+4\kappa^4}\right], \label{a10f} 
\end{eqnarray}
Figures~\ref{fig:three} and~\ref{fig:four} depict temporal profiles of
their intensities, for $\kappa<0.95$ (left graphs) and  $\kappa=1$ (right
graphs). 
Like in the previous case the seven localized probe modes are dominated by
periodic wavetrain profiles, consisting of sequences of pulses of different
periodicities when $\kappa<1$. From the right graphs we find that contrary to
the previous case, three distinct fundamental soliton modes show up when the XPM
strength is higher namely single-pulse, double-pulse and tripple-pulse periodic
wavetrains. Their analytical expressions are derived
from~(\ref{a10a})-(\ref{a10f}) for $\kappa\rightarrow 1$ and are explicitely
given by: \\
\begin{eqnarray}
u_1(\tau)&\propto& \left[2 - 5\,sech^2\left(\tau -
\tau_0\right)\right]\,tanh\left(\tau - \tau_0\right), \nonumber \\ 
k_1&=& k_{\lambda}, \label{a11a} 
\end{eqnarray}
\begin{eqnarray}
u_2(\tau)&\propto& \left[4 - 5\,sech^2\left(\tau -
\tau_0\right)\right]\,sech\left(\tau - \tau_0\right), \nonumber \\ 
k_2&=& k_{\lambda} + \frac{Q^2D}{2}, \label{a11c} 
\end{eqnarray}
\begin{eqnarray}
u_3(\tau)&\propto& \left[4 - 5\,sech^2\left(\tau -
\tau_0\right)\right]\,sech\left(\tau - \tau_0\right), \nonumber \\ 
k_3&=& k_{\lambda} + \frac{Q^2D}{2}, \label{a11b} 
\end{eqnarray}
\begin{eqnarray}
u_4(\tau)&\propto& sech^2\left(\tau - \tau_0\right)\,tanh\left(\tau -
\tau_0\right), \nonumber \\ 
k_4&=& k_{\lambda} + 2Q^2D, \label{a11e} 
\end{eqnarray}
\begin{eqnarray}
u_5(\tau)&\propto& sech^2\left(\tau - \tau_0\right)\,tanh\left(\tau -
\tau_0\right), \nonumber \\ 
k_5&=& k_{\lambda} + 2Q^2D, \label{a11d} 
\end{eqnarray}
\begin{eqnarray}
u_6(\tau)&\propto& sech^3\left(\tau - \tau_0\right), \nonumber \\ 
k_6&=& k_{\lambda} + \frac{9Q^2D}{2}, \label{a11g} 
\end{eqnarray}
\begin{eqnarray}
u_7(\tau)&\propto& sech^3\left(\tau - \tau_0\right), \nonumber \\ 
k_7&=& k_{\lambda} + \frac{9Q^2D}{2}. \label{a11f} 
\end{eqnarray}
Figure~\ref{fig:five} displays the spectrum of the seven localized soliton
modes, more precisely here are plotted their modulation wavectors versus
$\kappa$.
 Like the first case studied in the previous section, the modulation wavectors
decrease from a high-momentum excitation regime where the seven probe soliton
modes collapse into degenerate linear waves, to a low-momentum excitation regime
where they reduce to their characteristic fundamental soliton modes. Also
remarkable are mode degeneracies, in this case we notice that the number of
degenerate modes is higher and is expected to increase with increasing value of
the ratio of the XPM to the SPM coupling strengths. 
\section{conclusion}
The induced-phase-modulation (or cross-phase-modulation)
process has been predicted two decades ago~\cite{manas1,manas2,rand2,rand3}, and
experimentally observed~\cite{fuente1,fuente2} to occur when a low-power signal
(either an harmonic cw signal or a weakly nonlinear signal) couples to a
high-intensity pump signal, the later one providing an optical waveguiding
structure and reshaping channel to the first for substantial increase of its
lifetime. However while this process is rather well understood for
single-soliton pumps and in the case of relatively weak cross-phase-modulation
processes, almost no insight exists about the context of pump signals consisting
of periodic networks of temporal or spatial solitons~\cite{dika1}. The main
objective of this work was to point out the great variety of spectral properties
of pure harmonic cw probes trapped by a pump field consisting of a periodic
wavetrain of optical pulses, when different orders of magnitudes of the
cross-phase-modulation coupling are considered for the same
self-phase-modulation coupling strength. The principal motivation for carrying
out such analysis is the broad range of technological applications behind the
possibility to combine, both quantitatively and qualitatively, multi-color
soliton signals generated from on single pump field. On the other hand, the very
large diversity of physical properties expected from optical materials that can
be fabricated via current technological means, makes it hard to think of
physical contexts where cross-phase-modulation processes will always be far
weaker than the intrinsic nonlinearity of pump fields. From these last
standpoints the present study provides new interesting insight onto the general
physics of pump-probe systems. In particular the increasingly large number and
variety of possible soliton modes induced by the soliton pump for relatively
strong cross-phase-modulation proceses, offers new opportunity to create
multi-component vector solitons from optical low-power optical fields via
light-induced waveguiding phenomena, thus enriching current designs of
engineered (i.e. artificially created) soliton networks from reconfigurable
non-soliton signals by means of all-optical techniques. To end, we wish to
underline the richness of the fundamental modes emerging from the present study
with increasing cross-phase-modulation coupling strength, for a fixed strength
of the self-phase-modulation effect. While some of these modes have already been
predicted~\cite{manas2, rand2}, in the present and our recent~\cite{dika1} works
their spectral properties, conditions of existence as well as stability have
been formally established. The key insight about these fundamental modes is that
they put into play multi-pulse structures that are generated from the same pump,
but with different modulation wavectors and hence different instantaneous
wavelengths.
\section*{acknowledgments}
Work done in part at the Abdus Salam International Centre for Theoretical
Physics (ICTP) Trieste Italy. The author wishes to thank M. Marsili and M. Kiselev for their warm hospitality. \\


\section*{References}
\begin{thebibliography}{10}
\bibitem{manas1}Manassah J T 1990 {\it Ultrafast solitary waves sustained
through induced phase modulation by a copropagating pump} {\it Opt. Lett.} 15
670-672.
\bibitem{manas2}Manassah J T 1991 {\it Induced waveguiding effects in a
two-dimensional nonlinear medium} {\it Opt. Lett.} 16 587-589.
\bibitem{fuente1}De La Fuente R, Barthelemy A and Froehly C 1991 {\it
Spatial-soliton-induced guided waves in a homogeneous nonlinear Kerr medium}
{\it Opt. Lett.} 16 793-795.
\bibitem{fuente2}de la Fuente R and Barthelemy A 1992 {\it Spatial
soliton-induced guiding by cross-phase modulation} {\it IEEE J. Quantum
Electron.} 28 547-554.
\bibitem{stolen}Tomlinson W J, Stolen R H and Johnson A M 1985 {\it Optical wave
breaking of pulses in nonlinear optical fibers} {\it Opt. Lett.} 10 457-459.
\bibitem{agarwal}Agrawal G P, Baldeck P L and Alfano R R 1989 {\it Optical wave
breaking and pulse compression due to cross-phase modulation in optical fibers}
{\it Opt. Lett.} 14 137-139.
\bibitem{elena}Ostrovskaya E A, Kivshar Yu S, Skryabin D V and Firth W J 1999
{\it Stability of multihump optical solitons} \PRL 83 296-299.
\bibitem{malom}Shipulin A, Onishchukov G and Malomed B A 1997 {\it Suppression
of soliton jitter by a copropagating support structure} \JOSA B14 3393-3402.
\bibitem{rand2}Steiglitz K and Rand D 2009 {\it Photon trapping and transfer
with solitons} \PR A79 0218021-4(R).
\bibitem{rand3}Steiglitz K 2010 {\it Soliton-guided phase shifter and beam
splitter} \PR A81 0338351-5.
\bibitem{6}Eugenieva E D, Efremidis N K and Christodoulides D N 2001 {\it
Design of switching junctions for two-dimensional discrete soliton networks}
{\it Opt. Lett.} 26 1978-1980.
\bibitem{8}Fleischer J W, Carmon T, Segev M, Efremidis N K and Christodoulides D
N 2003 {\it Observation of discrete solitons in optically induced real time
waveguide arrays} \PRL 90 0239021-4.
\bibitem{9}Fleischer J W, Segev M, Efremidis N K and Christodoulides D N 2003
{\it Observation of two-dimensional
discrete solitons in optically induced nonlinear photonic lattices} {\it Nature}
422 147-150.
\bibitem{10}Neshev D, Sukhorukov A, Kivshar Yu S, Ostrovskaya E and Krolikowski
W 2003 {\it Discrete solitons in light-induced index gratings} {\it Opt. Lett.}
28 710-712.
\bibitem{11}Martin H, Eugenieva E D, Chen Z and Christodoulides D N 2004 {\it
Discrete solitons and soliton-induced dislocations in partially-coherent
photonic lattices} \PRL 92 1239021-4.
\bibitem{adolfo}Cartaxo A V T 1999 {\it Cross-phase modulation in intensity
modulation-direct
detection WDM systems with multiple optical amplifiers and dispersion
compensators} {\it J. Lightwave Technol.} 17 178-190.
\bibitem{naumov}Naumov A N and Zheltikov A M 2000 {\it Cross-phase modulation in
short light pulses as a probe for Gas ionization dynamics: the influence of
group-delay walk-off effects} {\it Laser Physics} 10 923-926.  
\bibitem{dika1}Dikand\'e A M 2010 {\it Fundamental modes of a trapped probe
photon in optical fibers conveying periodic pulse trains} \PR A81 0136211-5.
\bibitem{chen}Chen Z and Martin H 2003 {\it Waveguides and waveguide arrays
formed by incoherent light in photorefractive materials} {\it Optical Materials}
23 235-241.
\bibitem{mol1}Islam M N, Mollenauer L F, Stolen R H, Simpson J R and Shang H T
1987 {\it Cross-phase modulation in optical fibers} {\it Opt. Lett.} 12 625-627.
\bibitem{shapiro1}Shapiro J H and Bondurant R S 2006 {\it Qubit degradation due
to cross-phase-modulation photon-number measurement} \PR A73 0223011-4.
\bibitem{shapiro2}Shapiro J H 2006 {\it Single-photon Kerr nonlinearities do not
help quantum computation} \PR A73 0623051-11.
\bibitem{segev4}Lan S, DelRe E, Chen Z, Shih M F and Segev M 1999 {\it
Directional coupler using soliton-induced waveguides} {\it Opt. Lett.} 24
475-477.
\bibitem{menyuk1}Golovchenko E A, Pilipetskii A N and Menyuk C R 1996 {\it
Minimum channel spacing in filtered soliton wavelength-division-multiplexing
transmission} {\it Opt. Lett.} 21 195-197. 
\bibitem{menyuk3}Wai P K A, Menyuk C R and Raghavan B 1996 {\it Wavelength
division multiplexing in an unfiltered soliton communication system} {\it J.
Lightwave Technol.} 14 1449-1454.
\bibitem{duce}Del Duce A, Killey R I and Bayvel P 2004 {\it Comparison of
nonlinear pulse interactions in 160-Gb/s quasi-linear and dispersion managed
soliton systems} {\it J. Lightwave Technol.} 22 1263-1271.
\bibitem{hansen}Hansen E R 1975 \emph{A Table of Series and Products},
Prentice-Hill Inc, Englewood Cliffs(N. J.).
\bibitem{magnus}Magnus W, Oberhettinger F and Tricomi F G 1953 \emph{Handbook of
Transcendental Functions\/} (McGraw Hill, New York)
\bibitem{karta3}Aleshkevich V, Kartashov Y and Vysloukh V 2001 {\it
Self-frequency shift of cnoidal waves in a medium with delayed nonlinear
response} \JOSA B18 1127-1136.
\bibitem{karta5}Kartashov Y, Aleshkevich V, Vysloukh V, Egorov A A and Zelenina
A S 2003 {\it Stability analysis of $(1+1)$-dimensional cnoidal waves in media
with cubic nonlinearity} \PR E67 0366131-11.
\bibitem{dika2}Dikand\'e A M 1999 {\it Bound States in One-dimensional
Klein-Gordon systems admitting periodic-kink soliton excitations} \PS 60
291-293.
\bibitem{ars}Arscott F M 1964, \emph{Periodic Differential equations: An
introduction to Mathieu, Lam\'e and Allied Functions\/} (Pergamon Press LTD)
\end{thebibliography}
\end{document}